\documentstyle[preprint,aps,prb,epsf]{revtex}
\begin{document}
\preprint
\widetext
\title{Optical conductivity of the infinite-dimensional Hubbard model}
\author{M.\ Jarrell$^1$, J.\ K.\ Freericks$^2$ and Th.\ Pruschke$^3$}
\address{$^1$Department of Physics, University of Cincinnati, Cincinnati,
Ohio\\
$^2$Department of Physics, Georgetown University, Washington, DC 20057\\
$^3$Institut f\"ur Theoretische Physik, Universit\"at Regensburg,
93040 Regensburg, Germany\\
}
\date{\today}
\maketitle
\widetext
\begin{abstract}
	A Monte Carlo-maximum entropy calculation of the optical conductivity
of the infinite-dimensional Hubbard model is presented.  We show that the
optical conductivity displays the anomalies found in the cuprate
superconductors, including a Drude width which grows linearly with
temperature, a Drude weight which grows linearly with doping, and a
temperature and doping-dependent mid-IR peak. These anomalies arise as a
consequence of the dynamical generation of a quasiparticle band at the
Fermi energy as $T\rightarrow 0$,  and are a generic property of the
strongly correlated Hubbard model in all dimensions greater than one.
\end{abstract}
\renewcommand{\thefootnote}{\copyright}
\footnotetext{ 1994 by the authors.  Reproduction of this article by any means
is permitted for non-commercial purposes.}
\renewcommand{\thefootnote}{\alpha{footnote}}

\pacs{Principle PACS number XXXXX. Secondary PACS numbers XXXX}

\section{Introduction}
The discovery of high-T$_c$ superconductors based on CuO-compounds
\cite{bedmull} has led to a large amount of theoretical work about the
peculiar properties of these materials. A major effort has focussed on the
normal-state properties of these compounds.
This research was largely
motivated and substantiated by experiments that revealed striking anomalies
in the normal-state properties \cite{rew_oxides}. Most prominent among these
are the linear resistivity, a linear NMR-relaxation of the Cu-spins, and a
Hall angle that behaves $\sim T^2$ over a rather wide temperature region.
Furthermore, the optical conductivity shows a Drude peak with a
width\cite{romero,thomas} $1/\tau\sim T$, consistent with the linearity of the
resistivity,
and a Drude weight which grows linearly with doping, consistent with the notion
of holes acting as the charge carriers.  In addition, there
is a pronounced temperature and doping dependent mid-IR peak at
frequencies above the Drude peak.

It was argued from the beginning \cite{anderson87} that most of these anomalous
properties can be explained by two special features appearing simultaneously
in these materials: (i) They are strongly correlated, i.e.\ their (effective)
local Coulomb interaction is comparable to or larger than the characteristic
kinetic energy of the relevant carriers; and (ii) they are highly anisotropic
with the electrons being in principle confined to the CuO-planes characteristic
for these compounds.  Furthermore, it has been argued that the CuO planes
can be accurately described by a planar single-band Hubbard
model\cite{fczhang}.  The great interest in this class of materials has led to
a number of new theoretical conjectures, that, although based on the
assumption of strongly correlated carriers, focussed mainly on the 2D-character
of the CuO-planes\cite{pwa_rvb,solyom,pwa_lutt,ffnj}.

	In two earlier publications\cite{anom}, we found that
several of these anomalies can be understood from a Kondo-like effect in
the infinite-dimensional Hubbard model.
In particular, the density of states develops a sharp peak at the Fermi
surface as the temperature is lowered.  The development of this quasiparticle
peak coincides with the screening of the effective local moments and anomalies
in the transport. For example, the resistivity of the model displays
a distinct linear in $T$ behavior with a slope that increases in inverse
proportion to
the doping $\delta$, consistent with experiment\cite{christoph}.  The
NMR relaxation rate $1/T_1$ displays a pronounced linear in $T$
region, with a slope
that increases with doping, also consistent with experiment\cite{rew_oxides}.
Finally, the qualitative features of the Hall resistivity are
consistent with experiment\cite{rew_oxides}, including a
Hall angle that increases quadratically with the temperature.
Recently, a sharp Kondo-like peak at the Fermi surface has also been seen in
the
two-dimensional Hubbard model\cite{nejat} as the temperature is lowered.
{\em{This suggests that these anomalous normal-state properties are
intrinsic to the Hubbard model, independent of dimensionality!}}

In this contribution, we address the anomalous normal-state properties of the
Hubbard-model optical conductivity.  The optical conductivity $\sigma (\omega
)$
is an important probe of the excitations of a
strongly correlated system.  It measures the rate at which electron-hole
pairs are created by photons of frequency $\omega$. In a perfect
(translationally invariant) metal, photons couple only to electron-hole
pairs with vanishing momentum and energy; $\sigma (\omega )$ is proportional
to a Dirac delta function [$\sigma (\omega )=D_{perfect}\delta (\omega )$]
with Drude weight $D_{perfect}=\pi e^2n/m$ (we set $\hbar = 1$).
Electron-electron correlations modify this picture at zero temperature:
The charge and spin fluctuations induce a dynamic disorder to the lattice
potential which reduces the free-carrier Drude weight by the inverse of the
quasiparticle renormalization factor $Z$ ($D=Z^{-1}D_{perfect}$) and
transfers the remaining spectral weight to
a frequency-dependent component of $\sigma (\omega )$ reflecting
the incoherent charge and spin fluctuations; the total spectral weight is,
however, conserved.
Finite-temperature effects will broaden the zero-frequency delta function into
a Lorentzian and can modify both the quasiparticle renormalization
and the higher-frequency excitations.  This simple picture is further modified
when restriction is made to a single (or finite number) of electronic bands.
In this case, the total spectral weight can vary as a function of temperature
or interaction strength because the projection onto a restricted basis set
disregards all excitations to electronic bands that are higher in energy
than those that have been kept in the model.

If the on-site Coulomb repulsion $U$ is large enough, then the system will
be a Mott insulator at half-filling.  The single-particle density of states
$N(\omega )$ consists of two symmetric bands, separated by $U$ (called the
lower and upper Hubbard bands) with the Fermi level lying in the middle
of the gap.
The renormalization factor $Z$ diverges since there are no quasiparticles
at the Fermi level.  The optical conductivity will consist of a
charge-excitation peak centered at $\omega\approx U$, whose width is the
order of the bandwidth.  As the system is doped away from half filling,
a quasiparticle resonance appears at the Fermi energy as $T\rightarrow 0$.
The weight of the quasiparticle peak increases with doping.  In this case,
the optical conductivity will have a Drude peak (from the ``free''
quasiparticles at the Fermi energy) and a mid-IR band
(because of excitations between the lower
Hubbard band and the quasiparticle peak) in addition to the charge-excitation
peak. The Drude weight can be estimated as follows:  when $\delta$ holes are
doped into the half-filled band, the total electron concentration is
$(1-\delta)$, and the quasiparticle renormalization factor behaves
roughly as $Z\propto 1/\delta$, so the Drude weight is expected to be
\begin{equation}
D_{Drude}\propto Z^{-1}n \propto \delta(1-\delta)\quad ,
\label{eq: drude_approx}
\end{equation}
which increases linearly for small and large doping, and is peaked near
quarter filling $(n=0.5)$.  In addition, the weight of the mid-IR peak
should also behave qualitatively like Eq.~(\ref{eq: drude_approx}) because
it involves excitations between the quasiparticle peak and the lower Hubbard
band, which also should grow as the density of holes in the half-filled band.
The ``hole-like'' nature of the charge excitations
near half filling results from the strong Coulomb renormalizations
that create the Mott insulator at exactly half filling.

In this contribution, we investigate the optical conductivity of the Hubbard
model in infinite dimensions.  Our methodology is detailed in Section II, where
we discuss the formalism and calculational techniques.  Section III contains
our results, which include a Drude peak whose width $1/\tau$ grows linearly
with temperature, a mid-IR peak that becomes more visible at low temperatures
and dopings, and a charge-excitation peak that decreases with doping
and is weakly dependent upon $T$.  Comparison of our results with experiment
is presented in Section IV, and conclusions are given in Section V.

\section{Method}
Motivated by the observations of Anderson\cite{anderson87} and Zhang and
Rice\cite{fczhang} we will study the single-band Hubbard model\cite{hubbard}
in d-dimensions, with
\begin{equation}
H=-\frac{t^*}{2\sqrt{d}}\sum_{<ij>\,\sigma}c^\dagger_{i\sigma}c_{j\sigma}
+U\sum_in_{i\uparrow}n_{i\downarrow}\quad ,
\label{eq: hubmod}
\end{equation}
as a model for the CuO system.  Our notation is the following:
$c^{\dagger}_{i\sigma}$ is an electron creation operator for an electron
in a localized state at lattice site $i$ with spin $\sigma$; $n_{i\sigma}=
c^{\dagger}_{i\sigma}c_{i\sigma}$ is the corresponding electron number
operator; $t^*$ is the rescaled hopping matrix element;  and $U$ is the
Coulomb interaction strength.  We choose $t^*=1$ as a convenient unit
of energy.  This model may be solved in the limit of
high dimensions, using the observations of Metzner and Vollhardt \cite{mevoll}
that with increasing coordination number of the underlying lattice
the many-body renormalizations due to a two-particle interaction
like the Hubbard-$U$ in (\ref{eq: hubmod}) become purely local.

The solution of the model (\ref{eq: hubmod})
may be mapped onto the solution of purely local
correlated system coupled to an effective (self-consistently determined) bath
\cite{bramiel90,janis,kimkura,volljan,jarrell91,georges92}.
The quantum Monte Carlo (QMC) scheme based on
the work of Hirsch and Fye \cite{jarrell91,hifye,jarscal} has proven to be the
most effective
method for solving this strongly correlated local problem. Dynamical
properties of the model are then obtained using numerical analytic
continuation (employing maximum-entropy techniques)\cite{methods}.
This method requires a default model. To obtain the single-particle
density of states we use the finite-U non-crossing approximation
(NCA)\cite{nca}
result for the infinite-dimensional Hubbard model
as a default model at high temperatures, where it is essentially exact
\cite{pru_cox_ja}.  At
low temperatures, where the NCA fails, we use the numerically
continued result of the next higher
temperature as a default model.  The posterior probability of
the final result is employed to determine which default model should be used.
Generally, we find that the crossover temperature between using the
NCA default model and the higher-temperature continuation lies at
$T \approx 2T_0$, where $T_0$ is the Kondo-like
scale for this model\cite{anom}. Once the density of states is determined, the
self energy $\Sigma(\omega)$ may then be found by
(numerically) inverting the Fadeev function $w(z)$ in the relation
\begin{equation}
N(\omega)= {\rm{Re}} \left\{w[\omega+\mu-\Sigma(w)]\right\}/\sqrt{\pi}\,.
\end{equation}

With the knowledge of the one-particle self energy, one can calculate
transport quantities. For example, the optical conductivity can be calculated
exactly in the local approximation.  It is given by the
simple bubble only\cite{pru_cox_ja,khurana}, whose evaluation leads to
\begin{eqnarray}
\sigma_{xx}(\omega) &=& \frac{e^2\pi}{V}\int\limits_{-\infty}^\infty d\epsilon
\frac{f(\epsilon)-f(\epsilon+\omega)}{\omega}
\frac{1}{N} \sum_{\vec{k}\sigma}\left(\frac{\partial\epsilon_{\vec{k}}}
{\partial k_x}\right)^2
A(\epsilon_{\vec{k}},\epsilon)A(\epsilon_{\vec{k}},\epsilon+\omega)
\nonumber \\
&=&
\alpha\pi\int\limits_{-\infty}^\infty d\epsilon \frac{f(\epsilon)
-f(\epsilon+\omega)}{\omega}
\int\limits_{-\infty}^{\infty} dy \rho (y)A(y,\epsilon)A(y,\epsilon+\omega)\,,
\label{eq: optcond}
\end{eqnarray}
where $V$ is the lattice volume, $\alpha=e^2 /(Vd)$ and defines
the unit of the conductivity, the spectral weight satisfies
$A(\epsilon_k,\omega)=-\frac{1}{\pi}{\rm{Im}}\left[G(k,\omega)\right]$,
and the noninteracting density of states is $\rho (y)=\exp (-y^2 )/\sqrt{\pi}$.

As the temperature is lowered, the Hubbard model in infinite-d is found
to always be a Fermi liquid\cite{jarrell91,pru_cox_ja}, except
for the region of phase
space where it is magnetic\cite{mag_phase}.  A Fermi liquid is defined by a
self-energy that has the following structure:
\begin{eqnarray}
{\rm Re}\Sigma(\omega+i0^+)&=&{\rm Re}\Sigma(0)+\omega (1-Z)+O(\omega^2)
\quad ,\cr
{\rm Im}\Sigma(\omega+i0^+)&=&-\Gamma+O(\omega^2)\quad ,
\label{eq: flt}
\end{eqnarray}
with $\Gamma\propto T^2$ for temperatures $T\ll T_0$ the characteristic Fermi
temperature. [The Fermi temperature $T_0$ decreases to zero as half filling
is approached\cite{anom}, and the Fermi-liquid-theory form of
Eq.~(\ref{eq: flt}) still holds for moderate temperatures, with the only change
being $\Gamma\propto T$ for $T>T_0$.] The spectral weight, then assumes the
form
\begin{equation}
A(y,\omega) = {1\over\pi} {\Gamma\over \Gamma^2+(\omega Z+\epsilon_F-y)^2}+
A_{Inc}(y,\omega)\quad ,
\label{eq: spec_fun_flt}
\end{equation}
with the Fermi level defined by $\epsilon_F \equiv \mu - {\rm Re}\Sigma(0)$ and
$A_{Inc}(y,\omega)$ denoting the (rather structureless) incoherent
contributions
to the spectral function.  The spectral function includes a delta function
at zero temperature [$A(y,\omega) \rightarrow \delta(\omega Z+\epsilon_F-y)+
A_{Inc}(y,\omega)$] because the broadening $\Gamma$ vanishes in that limit.

If the Lorentzian form for the Fermi-liquid-theory spectral function
[Eq.~(\ref{eq: spec_fun_flt})] is substituted into the expression for the
optical conductivity found in Eq.~(\ref{eq: optcond}), and the temperature
satisfies $T\ll T_0\ll t^*$, then the optical conductivity becomes
\begin{equation}
{\sigma(\omega)\over\alpha}
={D\over \pi}{\tau\over 1+\omega^2\tau^2} + {\sigma_{Inc}(\omega)\over\alpha}
\quad ,
\label{eq: sigma_flt}
\end{equation}
with Drude weight $D=Z^{-1}\pi\rho(\epsilon_F)$, relaxation time
$\tau=1/(2\Gamma)\propto 1/T^2$,
and $\sigma_{Inc}(\omega)$ containing the contributions from the incoherent
pieces of the spectrum.

The noninteracting Drude weight for the
single-band model $(Z=1)$ satisfies $D_{non}=\pi\rho (\epsilon_F)$.
As the interaction $U$ is turned on in a
single-band model, the total integrated spectral weight is not conserved
but becomes $U$-dependent \cite{sum_rule}
because spectral weight that would be shifted to higher bands is ``lost''
in any single-band model. The total spectral weight satisfies
\begin{equation}
2\int_0^{\infty}\sigma(\omega )d\omega =
-\pi e^2 \left< T_x \right> \quad .
\label{eq: sumrule}
\end{equation}
where $\left< -T_x \right>$ is the kinetic energy per site, divided by
the number of lattice dimensions. It should be stressed that the modification
of the sum rule in Eq.~(\ref{eq: sumrule}) will produce some systematic
modifications to the behavior of the optical conductivity as a function
of temperature, interaction strength, and doping.  These systematic effects
must be kept in mind when one is comparing the results of a single-band
calculation to experiment.

The Drude weight $D$ for the interacting system, may also be determined
(at $T=0$) by
extrapolation of the Matsubara-frequency current-current correlation function
using the method proposed by Scalapino et al.\cite{scal_white_zhang}
($D$ measures the ``free'' quasiparticles in the system).
This method determines the Drude weight of a metal by
examining the asymptotic form of the current-current susceptibility in
the x-direction, $\Lambda_{xx}({\bf{q}},i\omega_n)$.  More specifically,
$D$ is given by
\begin{equation}
D=\lim_{T\rightarrow 0}\pi \Big [ e^2\left< -T_x \right> -
\Lambda_{xx}({\bf{q}}=0 , 2i\pi T)\Big ]\,,
\label{eq: scal_drude}
\end{equation}
where the limit $T\rightarrow 0$ is taken after first setting the
momentum transfer to zero (${\bf q}=0$).  Note that this latter method of
determining the Drude weight is a much better defined procedure than
trying to fit the optical conductivity to the generic form of
Eq.~(\ref{eq: sigma_flt}) because of the uncertainty left in trying to
fit $\sigma_{Inc}(\omega)$.

\section{Results}
\begin{figure}[t]
\epsfxsize=5.0in
\epsffile{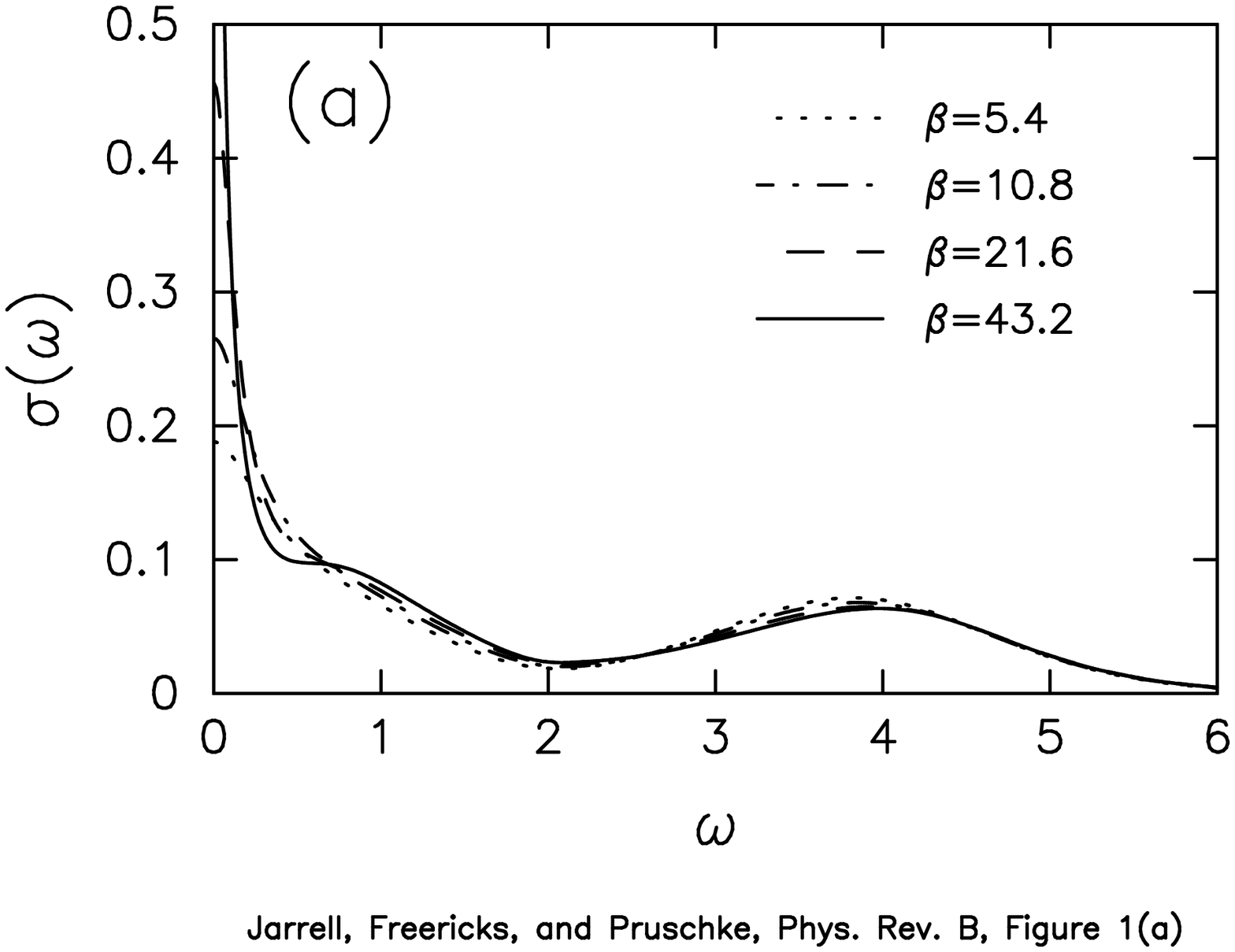}
\epsfxsize=5.0in
\epsffile{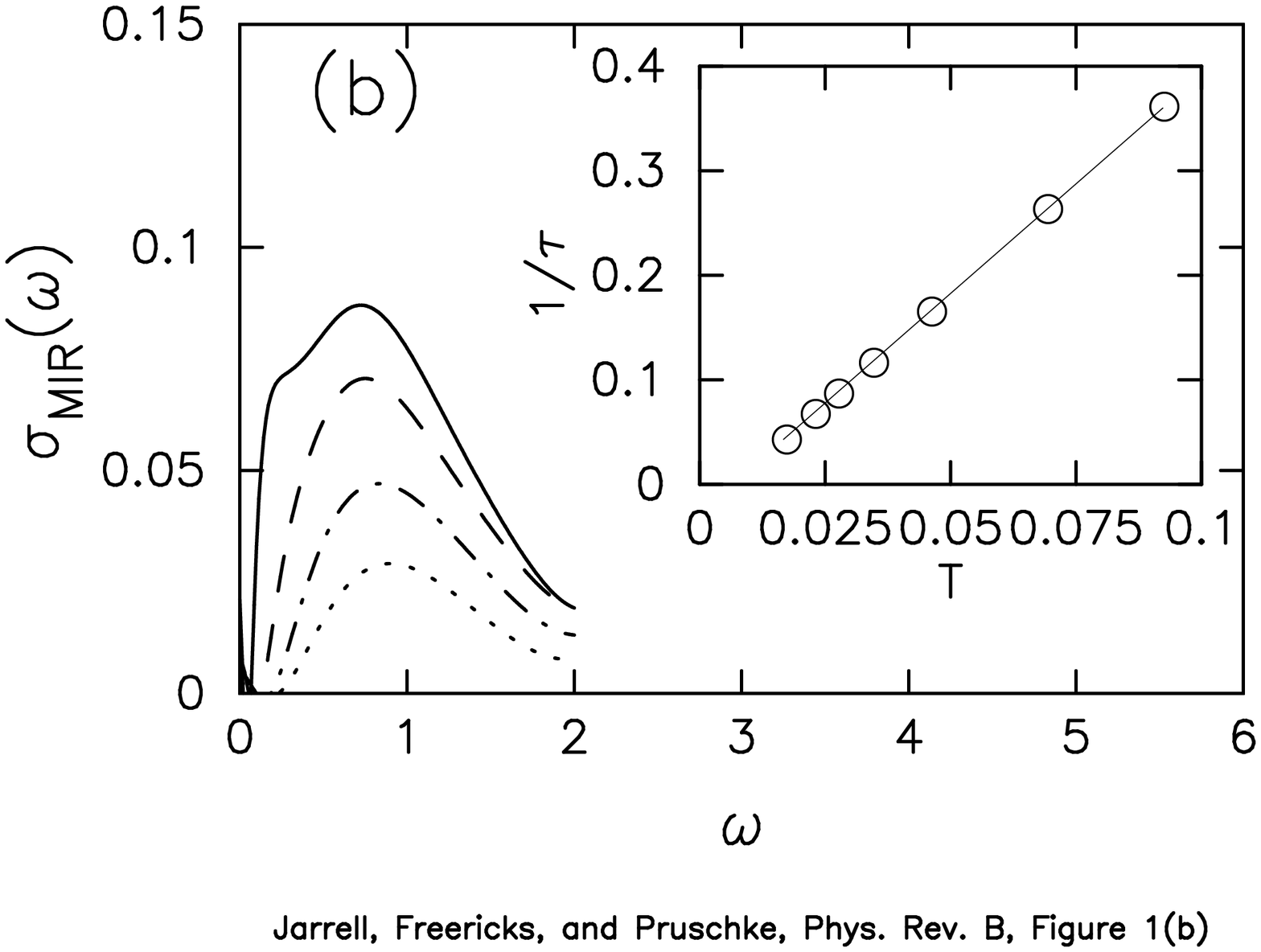}
\caption[]{\em {(a) Optical conductivity vs. $\omega$ for various temperatures
when $U=4$ and $\delta=0.068$ in units of $\alpha=e^2/(Vd)$.  Note that
at low temperatures, when the Kondo peak becomes pronounced in the DOS (see
Fig.~\ref{optical_dos}), a mid-IR feature begins to appear in $\sigma(\omega)$.
The mid-IR feature is isolated in (b) by fitting the low-frequency data with
Eq.~\ref{eq: opt_cond_fit},
and subtracting off the Drude part.  As the temperature is
lowered, the mid-IR peak becomes more pronounced and shifts to lower
temperatures.  Note that the double-peak structure in the mid-IR peak is
most likely due to the inaccurate fitting form for $\sigma_{MIR}(\omega )$.
As shown in the inset, the width ($1/\tau$) of the Drude peak is
found to increase roughly linearly with $T$.}}
\label{optical_mir}
\end{figure}

We present here our results for the optical conductivity of the single-band
Hubbard model in infinite dimensions with $U=4t^*$.
Figure~\ref{optical_mir}(a) shows the optical
conductivity obtained from Eq. (\ref{eq: optcond}) when $\delta=0.068$ for a
variety of temperatures.  One finds the Drude peak at $\omega=0$ developing
with decreasing temperature.  In addition, a shoulder develops adjacent
to the Drude peak at $\omega\approx 1$ which is strongly temperature
dependent and clearly visible only for the lowest temperatures.  The last
feature in $\sigma(\omega)$  is a roughly temperature-independent peak at
$\omega\approx U$. In order to compare our results to experiment, the three
features in $\sigma(\omega)$ are fit to a Lorentzian plus
(asymmetric) harmonic-oscillator forms for the higher-energy peaks
\begin{equation}
 {\sigma(\omega)\over\alpha}\approx \frac{D}{\pi}
\frac{\tau}{1+\omega^2\tau^2} + \frac{C_{MIR}}{\pi}
\frac{\omega^2\Gamma_{MIR}}{\omega^2\Gamma^2_{MIR}+(\omega^2-\omega_{MIR}^2)^2}
+ \frac{C_{C}}{\pi}
\frac{\omega^2\Gamma_{C}}{\omega^2\Gamma^2_{C}+(\omega^2-\omega_{C}^2)^2} ,
\label{eq: opt_cond_fit}
\end{equation}
with $\tau$ the relaxation time of the
quasiparticles, and the constants $C_{MIR}$, $\omega_{MIR}$, and $\Gamma_{MIR}$
($C_{C}$, $\omega_{C}$, and $\Gamma_{C}$)
the weight, center, and width, respectively of the mid-IR (charge-transfer)
peak.
The Drude width $1/\tau$ obtained from this fitting procedure is plotted in the
inset to Fig.~\ref{optical_mir}(b).  Note that for temperatures on the order
of $T_0$, it is
well approximated by a straight line (the line is a guide to the eye),
while for $T\ll T_0$ the Drude width must change its behavior to
$\tau\propto T^2$ according to the general properties of a Fermi liquid.  By
subtracting off the fit Drude portion from the optical conductivity,
we were able to isolate the mid-IR portion, as shown in
Fig.~\ref{optical_mir}(b).  Note that the ``double-peak'' structure emerging
in the mid-IR peak at the lowest temperature is an artifact of the fitting
procedure which is not perfect in extracting the parameters for the Drude peak.
The mid-IR peak is temperature dependent, growing in size and moving
to slightly lower frequencies as the temperature is lowered.

Comparison of the Drude weight $D$ determined by the fitting procedure in
Eq.~(\ref{eq: opt_cond_fit}) and the independent method of calculation
in Eq.~(\ref{eq: scal_drude}) produces only qualitative agreement.  We
have tried more sophisticated fitting routines in which we fit the
Fermi-liquid-theory parameters ($Z$, $\Gamma$, and
$\epsilon_F$) in the spectral function of
Eq.~(\ref{eq: spec_fun_flt})
and then determine the Drude contribution to the optical conductivity
by employing the full expression in Eq.~(\ref{eq: optcond}), and the
fit improves, as does the comparison of the Drude weight, but the quantitative
agreement still has a systematic error on the order of $10-20\%$ (where
the fitting procedure always
overestimates the Drude weight), because of the simple
form chosen for the mid-IR peak.  Our conclusion is that the method of
Ref.~\onlinecite{scal_white_zhang} is superior to any {\it ad hoc} fitting
procedure in determining the low-temperature Drude weight.  Note, however,
that the charge-transfer peak
is uniquely determined by the above fitting procedure (in the sense that
the results are independent of what fitting procedure is
employed), indicating that the
harmonic-oscillator form is a reasonable approximation to that peak.

The existence of the Drude and charge-excitation peaks
have been reported previously \cite{pru_cox_ja}. These results were obtained
with the NCA and were thus restricted to temperatures $T\agt 2T_0$.
The mid-IR bump, however, becomes clearly
visible only for temperatures below this
scale.  We can presently resolve this feature because of the refined numerics
employed in the QMC and  the numerical analytic continuation at lower
temperatures.

\begin{figure}[t]
\epsfxsize=5.0in
\epsffile{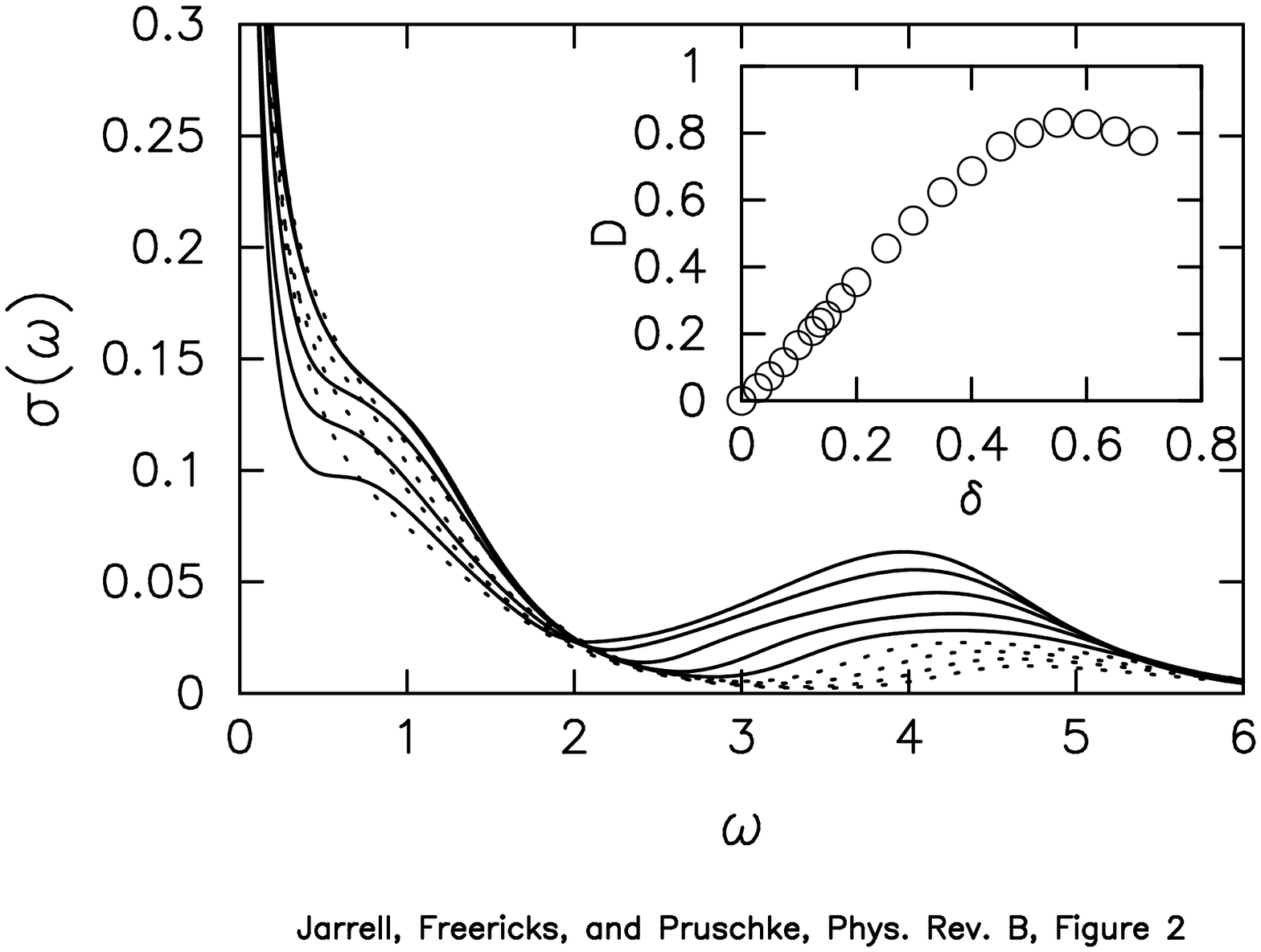}
\caption[]{{\em The doping dependence of the optical conductivity
when $U=4$ and $\beta=43.2$.  Note that for larger $\delta$ the mid-IR
and Drude peak begin to merge, so that the latter is less distinct.
The different
dopings are identified by their decreasing charge-transfer peaks, respectively
($\delta =0.068$, 0.0928, 0.1358, 0.1878, 0.2455, 0.3, 0.35, 0.4, and 0.45).
The solid lines correspond to the case where the optical conductivity
increases with doping in the lower-Hubbard-band region of $\omega < 2$
$(\delta < 0.25)$ and the dotted lines
correspond to the case  where $\sigma(\omega)$ decreases
with doping $(\delta > 0.25)$.
The inset shows the evolution of the Drude weight $D$ as a function of doping,
which is computed using Eq.~\ref{eq: scal_drude} at $\beta=20$.
For $\delta \alt 0.4$, the Drude weight increases linearly with doping.
}}
\label{optical_fill}
\end{figure}

	In Fig.~\ref{optical_fill}, the optical conductivity is plotted
as a function of doping for fixed temperature $\beta=43.2$.  The different
dopings are best identified by their decreasing charge-transfer peaks
($\delta =0.068$, 0.0928, 0.1358, 0.1878, 0.2455, 0.3, 0.35, 0.4, and 0.45).
The solid lines indicate the electron concentrations
where the optical conductivity
increases with doping in the lower-Hubbard-band region of $\omega < 2$
$(\delta < 0.25)$ and the dotted lines are where $\sigma(\omega)$ decreases
with doping in the same region $(\delta > 0.25)$.  One
can see that the mid-IR peak is strongly doping dependent, being
most distinct from the Drude peak at low dopings, and merging with
it as the doping and the width of the Drude peak increases.
In the inset, we show that the Drude weight $D$ increases linearly
with $\delta$. This latter result shows that the ``free'' carriers in a doped
Mott insulator are holes in the half-filled band.  As the doping is increased
further, the Drude weight eventually saturates, and then decreases
with doping, as the character of the ``free'' carriers changes from being
hole-like to being electron-like.

\begin{figure}[t]
\epsfxsize=5.0in
\epsffile{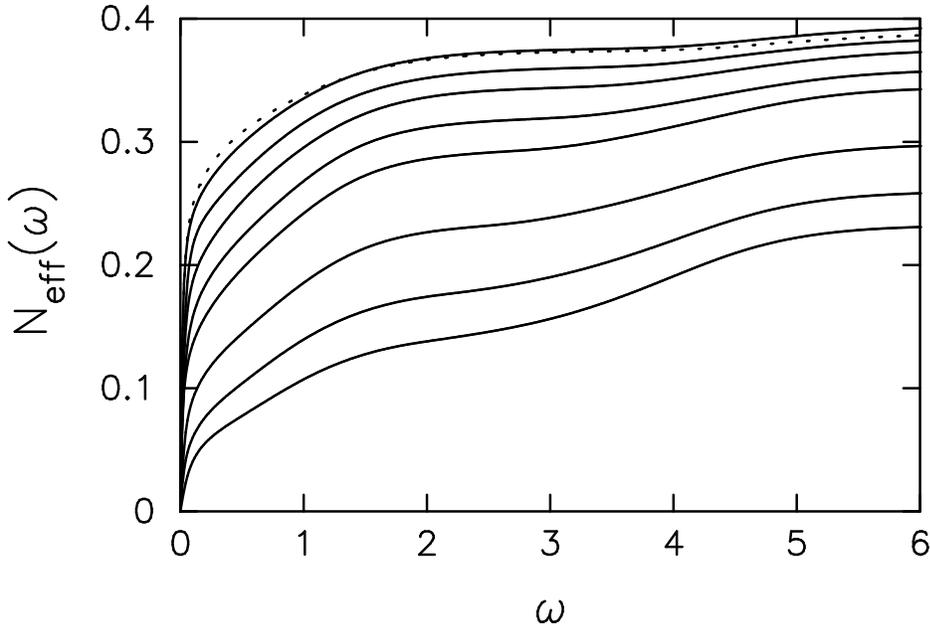}
\caption[]{\em Effective charge $N_{eff}(\omega )$ of the Hubbard model
when $U=4$ and $\beta=43.2$ as a
function of doping.  The same electron concentrations are plotted as in
Figure~\ref{optical_fill}.  The solid lines denote doping levels where the
effective charge increases with doping, and the dotted line corresponds
to the case where the effective charge decreases with doping ($\delta > 0.4$).}
\label{optical_int}
\end{figure}

The optical conductivity clearly displays an isobestic point, where
$\sigma(\omega)$ is independent of doping ($\omega_{IB}\approx
2)$.  The isobestic point marks the
boundary between the regions where the weight of the optical conductivity
increases as a function of doping $(\omega < \omega_{IB})$, and the regions
where the weight decreases with doping ($\omega > \omega_{IB}$).  However,
it is incorrect to assume that all of the spectral weight that lies below
the isobestic point was transferred from above.  An effective carrier number
$N_{eff}(\omega)$, defined by the integral of the optical conductivity
\begin{equation}
N_{eff}(\omega)={2\over \pi}
\int_0^{\omega} {\sigma(\Omega)\over\alpha} d\Omega\quad ,
\label{eq: n_eff}
\end{equation}
is plotted in Figure~\ref{optical_int} (the normalization is chosen here
to give the number of carriers for a model with a complete set of basis
states in order to make contact with experiment).
The integrated optical conductivity
increases rapidly with doping for small values of $\delta$, but then
increases more slowly, eventually saturating, and decreasing with doping
(at $\delta \approx 0.4$).  Once again, solid lines denote the electron
concentrations where $N_{eff}(\omega )$ increases with doping $(\delta < 0.4$)
and the dotted line is where $N_{eff}(\omega )$ decreases with doping
$(\delta > 0.4)$.
Note that there is no frequency where the integrated spectral weight is
independent of doping, which would be required for the above scenario
(of the Drude and mid-IR weight coming entirely from the charge-transfer
peak) to occur.

\begin{figure}[t]
\epsfxsize=5.0in
\epsffile{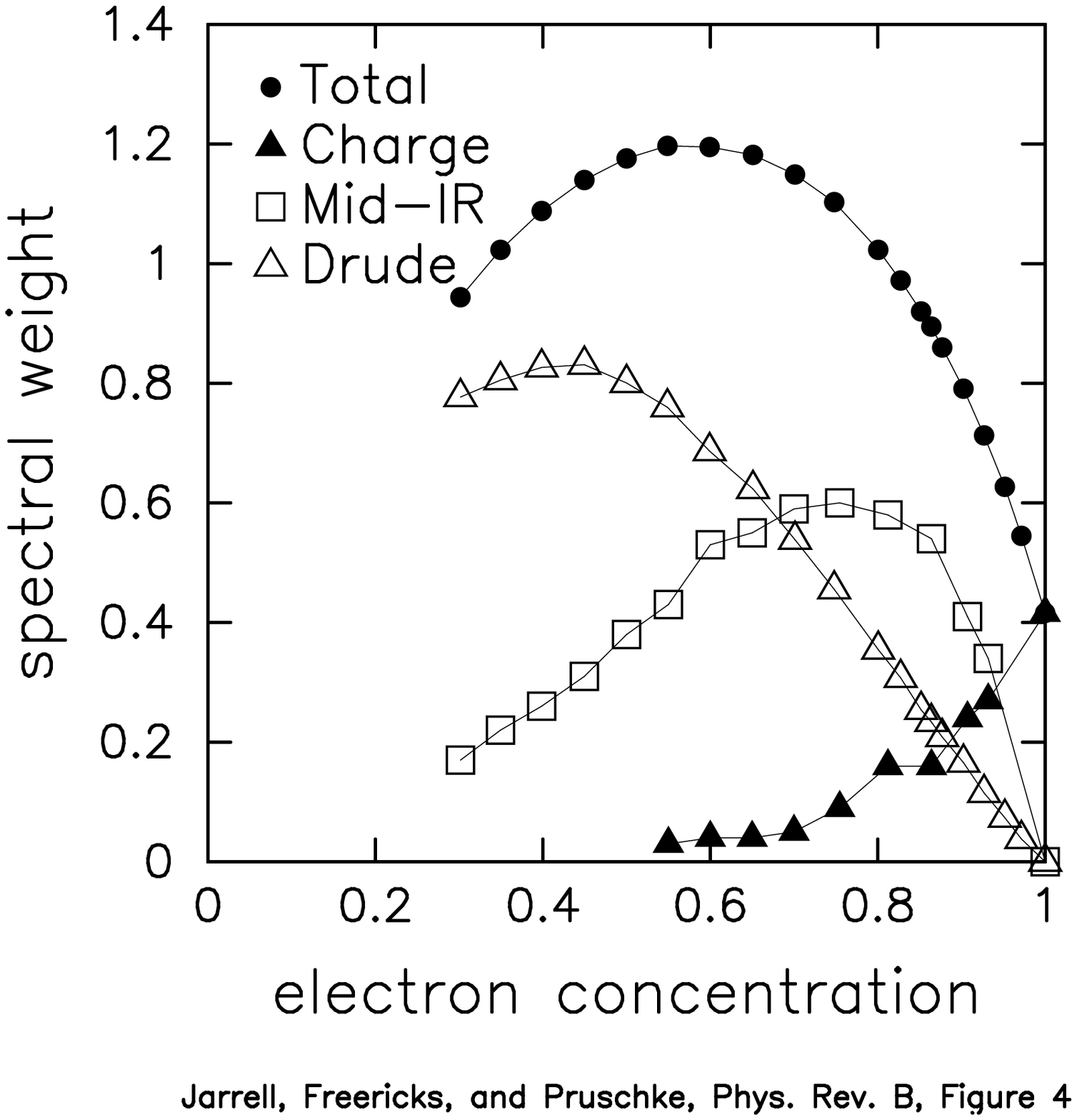}
\caption{\em Total spectral weight and its individual components for the
optical conductivity of the Hubbard model as a function of doping at $U=4$ and
$\beta = 43.2$.  The total spectral weight (solid dots) is broken up into
its component pieces (Drude [open triangles], mid-IR [open squares], and
charge-transfer [solid triangles]).  Note how the charge-transfer peak is
important only as the electron concentration nears half filling, and how the
mid-IR peak disappears as the electron correlations become smaller.}
\label{optical_weights}
\end{figure}

In addition, we can examine how the sum rule for the total spectral weight,
and for it's component pieces (Drude, mid-IR, and charge-transfer) evolve
as a function of doping at fixed low temperature (in fact, the following
analysis is an approximation to the zero-temperature behavior of the different
components of the spectral weight).  This is shown in
Figure~\ref{optical_weights}.  The total spectral weight is found from
Eq.~(\ref{eq: sumrule}) (which agrees with the integral of
the optical conductivity to within 3\%), the charge-transfer weight from
the fitting procedure in Eq.~(\ref{eq: opt_cond_fit}), the Drude weight
from Eq.~(\ref{eq: scal_drude}), and the weight of the mid-IR peak is
then determined by subtraction (the charge-transfer weight is neglected for
$n < 0.5$).  Note how the mid-IR peak initially increases linearly
with doping until $n\approx 0.75$ where it saturates and then decreases as the
system becomes less correlated, and how the system changes from a hole-like
metal to an electron-like metal at about quarter filling.  In addition, one can
easily see that the Drude and mid-IR weights both grow quite rapidly as the
system is doped away from half filling (the initial growth rate of the
mid-IR weight is about a factor of 2 faster than that of
the Drude weight). This fast growth of the Drude and mid-IR weights explains
why the optical conductivity for $\omega < \omega_{IB}$ increases so rapidly
with doping as shown in Fig.~\ref{optical_fill}.  These results are
qualitatively similar to those found in the strong-coupling limit of the
Hubbard
model in one dimension\cite{oles}, indicating that the optical properties
of the Hubbard model are not strongly dependent upon the dimensionality
(with the exception that there is no mid-IR peak in one dimension).

One can also inquire into the temperature dependence of the Drude, mid-IR, and
charge-transfer peaks at fixed doping.  The analysis, in this case, relies
on trying to fit the optical conductivity to the form in
Eq.\ \ref{eq: opt_cond_fit}, as a function of temperature.  Such a procedure
is ill-defined, because the Drude and mid-IR peaks merge at high temperatures,
and it is difficult to separate them into their component pieces, without
explicit knowledge of the proper fitting forms. However, the qualitative
physics can be discussed:  We find that as the temperature is raised, that
spectral weight is lost from the system (i. e., the magnitude of the
expectation value of the kinetic-energy operator decreases as $T$ increases),
that the Drude and charge-transfer weights slightly increase, and that the
mid-IR weight is sharply reduced.  In addition, spectral weight rapidly
shifts to higher frequency, since the Drude width (which is easy to determine
by the fitting procedure) increases linearly with $T$ for temperatures
larger than $T_0$.

\begin{figure}[t]
\epsfxsize=5.0in
\epsffile{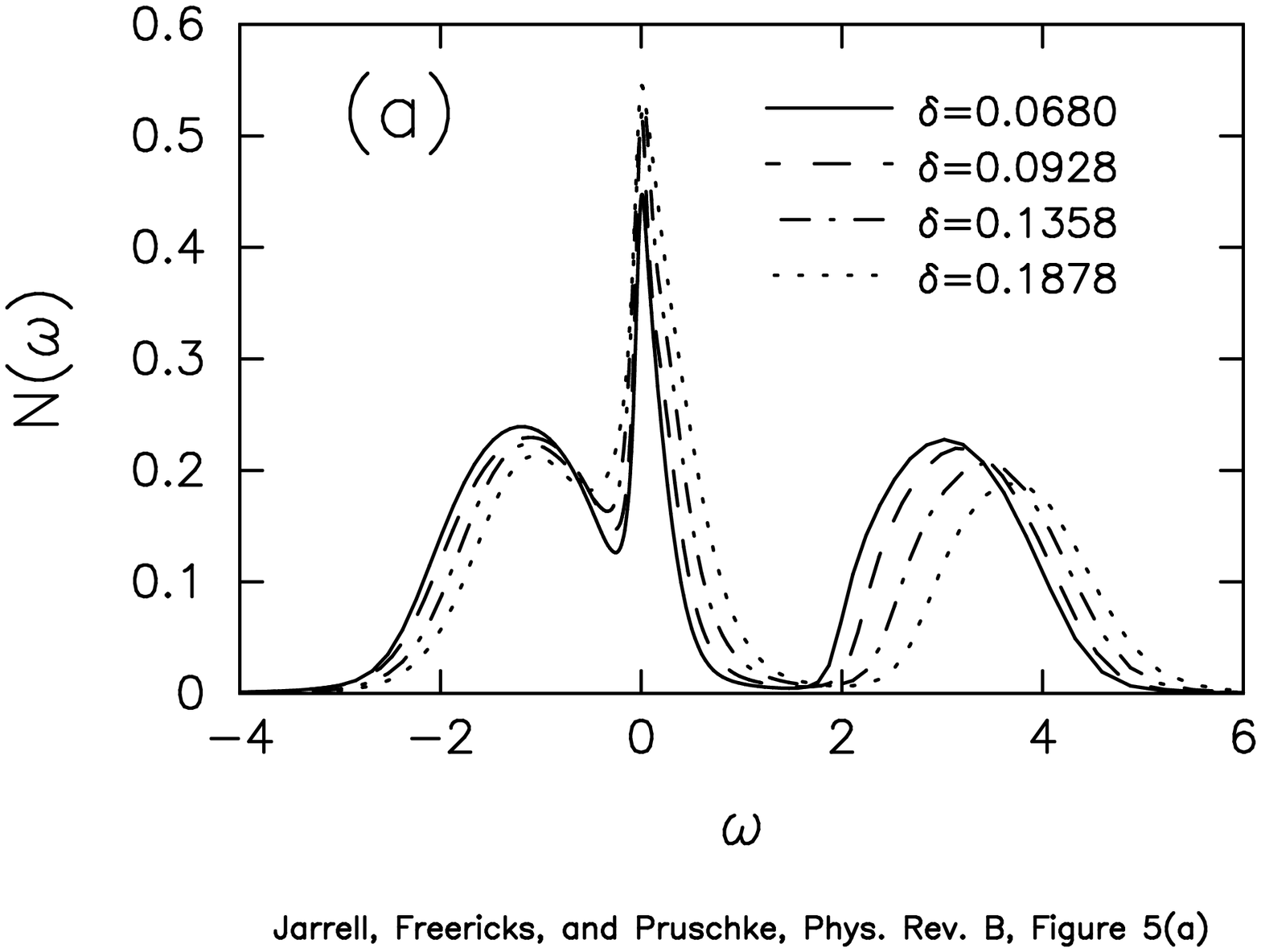}
\epsfxsize=5.0in
\epsffile{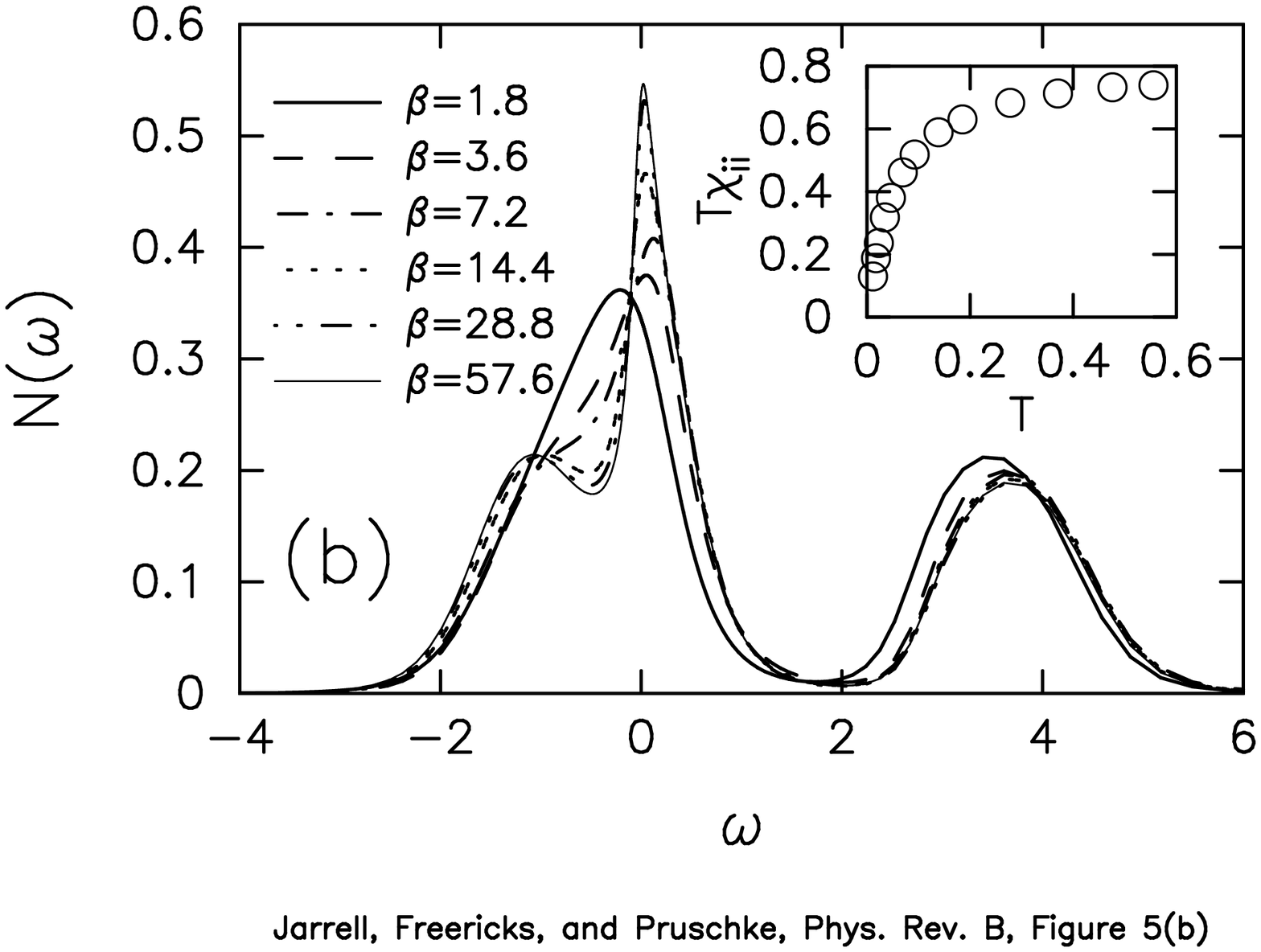}
\caption[]{{\em One-particle DOS of the Hubbard model.  In (a) the
DOS is plotted for different dopings $\delta=1-n$ at
$\beta=43.2$. In addition to the lower and upper Hubbard bands,
which are fairly doping independent, a resonance occurs at the chemical
potential that becomes broader with increasing $\delta$ and finally merges
with the lower Hubbard band.  In (b), the temperature
evolution of the density of states is plotted for
$U=4$ and $\delta=0.188$. As the temperature is lowered, a sharp
peak, distinct from the lower Hubbard band, develops at the Fermi surface.
As shown in the inset, the development of a sharp peak at the Fermi surface
is correlated with the reduction of the screened local moment and hence is
associated with resonant Kondo screening of the spins.}}
\label{optical_dos}
\end{figure}

The origin of all of these features in the optical conductivity can be
understood by studying the single-particle density of states, plotted as a
function of both temperature (a) and doping (b) in Fig.~\ref{optical_dos}.
Clearly the Drude peak results from the quasiparticle band which develops at
low temperatures.  We have shown previously that while this peak is developing
({\it i. e.} in the region $T > T_0$),
the scattering rate (as measured by the resistivity) increases in proportion
to the temperature\cite{anom}.  Thus, the Drude width also increases in
proportion to the temperature.  We attribute the mid-IR peak to
excitations from the lower Hubbard band to the quasiparticle band at the
Fermi energy.  Note that the quasiparticle resonance is only sharp and
distinct from the lower Hubbard band at low $T$ and small $\delta$.
Thus, the mid-IR peak will only be visible in the optical conductivity at
low $T$ and small $\delta$ as well.  On the other hand, the
upper peak at $\omega \approx 4$ is due to the charge excitations from
both the lower Hubbard band and the quasiparticle band to the upper
Hubbard band.  Since the upper Hubbard band is distinct from the
quasiparticle weight at the Fermi surface,  and since the weight in the
upper Hubbard band does not change as
significantly with doping, the upper peak in
$\sigma(\omega)$ has a  weaker temperature and doping dependence, but is
expected to disappear as the system becomes uncorrelated in the low-density
limit.

\section{Comparison to Experiment}

Many experimental measurements have been made of the optical conductivity
in doped cuprates.  Kramers-Kronig analysis is used to determine
$\sigma(\omega)$ from reflectivity measurements in\cite{uchida1}
La$_{2-x}$Sr$_x$CuO$_4$ (LSCO), in\cite{orenstein,thomas2}
YBa$_2$Cu$_3$O$_{6+x}$ (YBCO), in\cite{romero,romero2}
Bi$_2$Sr$_2$CaCu$_2$O$_x$
(BSCCO), and in\cite{uchida2,thomas2} Nd$_2$CuO$_{4-y}$ (NCO); reviews have
also
been published\cite{thomas}.  Photoinduced
absorption is also employed to measure the optical conductivity
in\cite{heeger} YBCO and in both\cite{young} LSCO and NCO.
It is interesting to compare the theoretical calculation
of the optical conductivity with these experimental results.

The experiments yield six trends for the cuprate superconductors:
1) The mid-IR peak maximum moves to lower frequency, and merges with the
Drude peak as the doping increases; its spectral weight grows very rapidly
with doping near half filling; 2) At a fixed value of the doping, spectral
weight rapidly moves to lower frequency as $T\rightarrow 0$, but the total
weight in the Drude and mid-IR peaks remains approximately constant; the
width of the Drude peak decreases linearly with $T$; 3) The
insulating (or undoped) phase has a charge-transfer gap; when doped, the
optical conductivity initially increases within the gap region, but eventually
saturates and then decreases with doping; 4) There
is an isobestic point or nearly isobestic behavior (in that the optical
conductivity is nearly independent of doping) at a frequency that is
approximately one half of the charge-transfer gap; 5) The effective charge
has a constant value with respect to doping near the charge-transfer edge;
and 6) more than one peak is observed within the mid-IR region.

Most of these trends are reproduced by our theoretical model.  In particular,
Fig.~\ref{optical_fill} illustrates how the mid-IR peak moves to lower
frequency
and joins the Drude peak (1), how the optical conductivity initially increases
with doping at low frequency, but then saturates and decreases (3), and how
there is an isobestic point (4).  Fig.~\ref{optical_mir}
shows how the spectral weight is transferred to lower frequencies
as the temperature is lowered, but we find that the total (Drude plus mid-IR)
spectral weight does not remain constant as in (2) because the temperature
dependence of the expectation value $<T_x>$ produces some temperature
dependence to the total Drude plus mid-IR spectral weight (it actually
decreases as $T$ increases).  Fig.~\ref{optical_int} does not display
the trend of point (5), possibly because the restriction to a single-band
model reduces the optical conductivity at higher frequencies to such an
extent that the effective charge must depend on the doping level for the
theoretical model.  We do not see the multiple peaks in the mid-IR region
of point (6).

These experimental
features are usually attributed\cite{polarons} to phonons or impurities present
in the system, but judging from our results, the low-energy feature may
also be due to excitations from the lower
Hubbard band to a dynamically generated quasi-particle band at the Fermi
energy (which is generated by Kondo-like screening of the moments). Naturally,
the single-band model fits the experminental data much better at lower
frequencies, where the single-band approximation is relevant, but fails
in reproducing some of the higher-energy trends found in the cuprates, because
of the neglected bands. A
comprehensive theory should include both the effects of strong electron
correlation (which reproduce the insulator at half filling, and give a
hole-like
character to the charge excitations near half filling) with the effect of
electron-phonon coupling (which are necessary to explain similar mid-IR
features
in nonsuperconducting perovskites).

\section{Conclusion}
A quantum
Monte Carlo and maximum entropy calculation of the optical conductivity
of the infinite-dimensional Hubbard model has been presented.
The Mott-insulating character of the ground state at
half filling drives many anomalous behaviors in the normal state near
half filling that are similar to those observed in the cuprate superconductors.
In particular the system is always a Fermi liquid away from half filling, but
the Fermi temperature vanishes, and the quasiparticle renormalization factor
diverges as half filling is approached.  As a result, the free carriers in the
system initially
have a hole-like character (that changes to an electron-like character
at approximately quarter filling).  The Drude width for these carriers grows
linearly with temperature for temperatures above $T_0$, the Drude weight
grows linearly with doping, and there is a doping and temperature dependent
mid-IR peak.  These anomalies arise naturally from the presence of a strongly
temperature-dependent quasiparticle peak, whose origin is a Kondo-like
screening of the magnetic moments, and which appears to occur in the Hubbard
model for all dimensions greater than 1.

The anomalous features in the experimentally measured optical conductivity
for the cuprates are usually attributed to either polarons or impurities.
However, any purely polaronic theory has difficulty in explaining the
magnetic insulating character of the ground state at half filling.  The
Hubbard model naturally describes such an insulating state, and appears to
also describe many of the anomalous features present in the experimental
data.  In light of this fact, it is worthwhile to try to incorporate both
the effects of strong electron correlation, and the electron-phonon interaction
into a comprehensive theory for the normal state of the cuprate materials.
Work along these lines is in progress.

\acknowledgments
We would like to acknowledge useful conversations with
W.\ Chung,
J.\ Keller,
Y.\ Kim,
D.\ Scalapino,
R.\ Scalettar,
D.\ Tanner,
and G.\ Thomas.
This work was supported by the National Science Foundation grant number
DMR-9107563, the NATO Collaborative Research Grant number CRG 931429 and
through the NSF NYI program.  In addition, we would like to thank the Ohio
Supercomputing Center, and the physics department of the Ohio State University
for providing computer facilities.

\epsfclipoff

\end{document}